%% file: 0_main.tex
\documentclass[sigconf,natbib=false]{acmart}

\AtBeginDocument{%
  }

\acmYear{2025}
\acmConference[SIGCSE TS 2025]{Proceedings of the 56th ACM Technical Symposium on Computer Science Education V. 1}{February 26-March 1, 2025}{Pittsburgh, PA, USA}
\acmBooktitle{Proceedings of the 56th ACM Technical Symposium on Computer Science Education V. 1 (SIGCSE TS 2025), February 26-March 1, 2025, Pittsburgh, PA, USA}
\acmDOI{10.1145/3641554.3701820}

\setcopyright{none}

\makeatletter
\renewcommand\footnoterule{%
  \kern-3\p@
  \hrule \@width \columnwidth \kern 2.6\p@}
\makeatother

\renewcommand{\footnotetextcopyrightpermission}[1]{%
  \footnotetext[0]{%
    \footnotesize
    This is the author's version of the work. It is posted here for your personal use. Not for redistribution. The definitive Version of Record was published in \textit{Proceedings of the 56th ACM Technical Symposium on Computer Science Education V. 1 (SIGCSE TS 2025), February 26--March 1, 2025, Pittsburgh, PA, USA}, \url{https://doi.org/10.1145/3641554.3701820}.%
    \vspace{5\baselineskip}%
  }%
}

\usepackage{balance}

\RequirePackage[
  datamodel=acmdatamodel,
  style=acmnumeric,
  ]{biblatex}

\begin{document}

\title{Enhancing Student Performance Prediction In CS1 Via In-Class Coding}



\author{Eric Hicks}
\email{elhicks@memphis.edu}

\affiliation{%
 \institution{The University of Memphis}
 \city{Memphis}
 \state{TN}
 \country{USA}
}

\author{Vinhthuy Phan}
\email{vphan@memphis.edu}

\affiliation{%
 \institution{The University of Memphis}
 \city{Memphis}
 \state{TN}
 \country{USA}
}

\author{Kriangsiri Malasri}
\email{kmalasri@memphis.edu}

\affiliation{%
 \institution{The University of Memphis}
 \city{Memphis}
 \state{TN}
 \country{USA}
}

\renewcommand{\shortauthors}{Eric Hicks, Vinhthuy Phan, and Kriangsiri Malasri}

\begin{abstract}
Computer science’s increased recognition as a prominent field of study has attracted students with diverse academic backgrounds.
This has significantly increased the already high failure rates in introductory courses.
To address this challenge, it is essential to identify struggling students early on.
Incorporating in-class coding exercises in these courses not only offers additional practice opportunities to students but may also reveal their abilities and help teachers identify those in need of assistance.
In this work, we seek to determine the extent to which the practice of using in-class coding exercises enhances the ability to predict student performance, especially early in the semester.
Based on data obtained in a CS1 course taught at a mid-size American university, we found that in-class exercises could improve the prediction of students' eventual performance.
In particular, we found relatively accurately predictions as early as academic weeks 3 through 5, making it possible to devise early intervention strategies.
This work can benefit future studies on the impact of in-class exercises as well as  intervention strategies throughout the semester.
\end{abstract}



\maketitle

\input{1_intro.tex}
\input{2_related_work.tex}
\input{3_method.tex}
\input{4_result.tex}
\input{6_conclusion.tex}

\balance
\printbibliography

\appendix
\end{document}

%% file: 1_intro.tex
\section{Introduction}
\label{section:intro}
CS1 is known to have a particularly high failure and a dropout rate of more than 30\% \cite{liao19behaviors, liao19value, porter14, robins10}. The constant increase of enrollment in these courses means that more students are falling behind and coupled with the tight integration of topics, staying behind \cite{porter2014importance, becker16}.  This has made it especially important to predict student performance early on, as falling behind in the early stages of any course can lead to poor performance and dropping out \cite{vandamme2007predicting, petersen2016revisiting, liao19behaviors}. These factors have led efforts to predict student outcomes earlier and earlier in a course \cite{mao19} so that instructors have the most time to intervene. It has been shown that test results from the 3rd week are predictive of final scores in CS1 \cite{robins2003learning, porter14}. By knowing which students need help early on, instructors can utilize more strategies to help struggling students change their study habits and correct mistakes \cite{liao2016lightweight}, which has also been shown to help them improve in later, more advanced courses \cite{bergin2006predicting}.

Much effort has been focused on predicting student success with traditional assignments \cite{liao19robust, mao20, castro2017evaluating, romero2008data}, however there is often a large gap between when a student does the work and when their performance on it can be used.
This delay allows for bad habits and misconceptions to grow in students before they can be identified and receive help.
To help improve overall student success in CS1, instructors have implemented active learning activities such as clicker use and in-class coding exercises \cite{prince2004does}.
By using these in-class assignments, which often have in-class feedback, students who need help can be identified almost instantly \cite{hicks22}.
Further, several methods have been utilized to collect in-class coding information unobtrusively for subsequent analysis \cite{Phan2018, lyulina21, kinnebrew13}.

We hypothesize that the use of this in-class coding data can enhance our ability to predict student performance throughout the semester.
We think that because in-class exercises are immediately given when students are taught specific concepts and skills, their evaluations can be indicative of understanding and learning.
Further, in-classes exercises are given more often than traditional assignments and assessed almost immediately before classes end, giving us the advantage of having more readily available data for analysis.
Understanding how in-class exercises can help improve our prediction of student performance should empower researchers and instructors to come up with timely interventions.
To this end, we are interested in two research questions on the impact of in-class exercises on predicting student performance.  Further, we are also interested in knowing if an instructor's curved grading scheme, can alter our findings.
These research questions are:\\

\noindent
{\bf RQ1: Can in-class exercises enhance the ability to predict students' eventual performance?} As Active Learning literature has already shown the benefits of using in-class exercises to help students learn coupled with their more frequent and plentiful use, we hypothesize that including these features should improve models predicting student performance.\\

\noindent
{\bf RQ2: To what extent does a curved final course grade impact prediction ability?} Curved grading is a controversial, yet widely practiced method \cite{kulick2008impact, dubey2010grading, tan2020students, czibor2014does, schinske2014teaching}. To this end it is important to determine the effect this has on the model, both to improve our ability to help students as well as to make our results fully applicable to other instructors, courses, and universities.\\

%% file: 2_related_work.tex
\section{Related Work}
\label{section:related_work}
Recent work in computer science education has employed advances in machine learning methods to study, model, and make predictions of student performance \cite{liao19robust, mao19, wang20, liao19value, mao20}.
Liao et. al \cite{liao19value} used grades in prerequisite computer science courses, assignments, online quizzes, and clicker data from five different CS courses to build generalized linear models to predict student performance in a subsequent semester.
In a related study, Liao et. al \cite{liao19robust} built support vector machines (SVM) models based on clicker data collected from Peer Instruction activities to  identify students at risk of failure so that these students could receive support and take corrective action.
Mao et. al \cite{mao19} used data from iSnap \cite{price2017isnap} to build models to predict whether a student would have difficulty making progress within the next five minutes. These models are composed of Recent Temporal Patterns in conjunction with SVM and Logistic Regression.
To model the dynamics of student knowledge state in continuous time, Mao et. al \cite{mao20} used a type of deep learning model known as Time-aware Long-Short Term Memory to capture temporal characteristics during iSnap coding sessions.
Wang et. al \cite{wang20} used SVM with a linear kernel and regularization to predict complex behaviors of students while they designed Scratch games.
In general, these machine learning methods model the relationships between a \emph{target variable} and a set of \emph{features}, which are selected or extracted from modeling data.

\subsection{Modeling Data And Target Variables}
To predict student performance in coding courses, researchers have used different data sources including those generated from clickers \cite{porter14, liao19value, liao19robust}, compilers \cite{becker16}, and student code written in block-based languages \cite{mao19, gao21, mao20, zhi2018reducing}.
To define target variables from this data, it is necessary to specify thresholds that reflect how instructors normally assess performance.

Liao et al. have separated data into higher and lower performing groups at the 40\% \cite{liao19robust, liao19value} or 50\% mark \cite{liao19behaviors}.  Other researchers have split data into three groups of perfect thirds represented by student grades \cite{mao20} or into categories such as \emph{needs help, good work}, and \emph{great work} \cite{porter14}.

\subsection{Features And Feature Engineering}
Models are trained on features that are selected from raw data by either domain experts or automated methods. These features are hypothesized to influence the outcome (or target variable) in some ways.

Early work utilized features that existed prior to when students enrolled in the course such as past course grades, GPA, and prior understanding of course materials \cite{gao21, liao19robust, liao19value}, programming self-efficacy \cite{bergin2005influence, bergin2006predicting, quille2018programming}, and mathematical ability \cite{bergin2006predicting, bergin2005programming,  wilson2001contributing, quille2018programming}.

These features can be difficult for instructors to collect and do not directly represent how students perform in the current course. As such, it is not easy for instructors to construct models and use them to predict performance for a specific course.

To address this, several authors have used student performance throughout the course of a semester as features to predict final grades or relative final performance in a course \cite{porter14, liao19robust, liao19value, mao20, castro2017evaluating, romero2008data}. Liao et al. further attempted to use current semester performance to predict performance in a subsequent course in the next semester \cite{liao2016lightweight}. However, this type of modeling does not show how students perform throughout the course of a semester and therefore may not be effectively used for intervention.

To get more information from the problem-solving process, researchers have analyzed student behaviors throughout the entire course \cite{liao19behaviors} and student code to extract useful features.
Features derived from student code include abstract syntax trees \cite{wang20}, encoded patterns that show how students interact with the programming environment \cite{emerson19, gao21}, and temporal patterns that show how students problem-solve through time \cite{mao19, mao20}.
These approaches make use of assignments that often take days to grade and therefore might not generate data early enough for timely intervention after modeling.
To combat this delay, Liao et al. \cite{liao19value, liao19robust} and Porter et al. \cite{porter14} began using in-class activities such as clicker data to achieve great predictive outcomes, while Gao et al. \cite{gao21}, Mao et al. \cite{mao19, mao20}, and Zhi et al. \cite{zhi2018reducing} began using coding data written in block-based languages as features with similarly positive results.

Given that short programming assignments in high-level languages have resulted in positive learning outcomes \cite{kulkarni2015peerstudio, allen19, prince2004does, bjork2011making}, it is interesting to find out if short in-class coding exercises in a high-level language can be used as features for modeling student performance throughout the course of a semester.

\subsection{Curved Grading}
Curved grading is a method where students' grades are assigned based on their performance relative to their peers, for example by adjusting the grading scale based on the highest scores.
Curved grading has been shown to encourage students\cite{dubey2010grading} and improve student performance \cite{czibor2014does}.  It has also been shown to discourage students \cite{tan2020students} and worsen student performance \cite{schinske2014teaching}.  

These conflicting results have made curved grading controversial, but it is still commonly used \cite{kulick2008impact, dubey2010grading, tan2020students, czibor2014does, schinske2014teaching}.
As such, it can be helpful to consider modeling and predicting student performance in the presence of curved grading.

%% file: 3_method.tex
\section{Method}
\label{section:method}
We investigated the viability of using in-class coding exercises as features to model and predict student performance on lab finals, final exams, and final grades.
Additionally, we compared and contrasted these features with and against more traditional features, such as homework and exams.
We further explored the use of various grade scale target sets to achieve the best predictive outcomes.

\subsection{Participants}

This study involved historical data of approximately 300 undergraduate students in seven sections of CS1 over the Spring, Summer, and Fall semesters of 2019.  Section sizes ranged from 22 to 60, with an average size of 43 students.  The Spring and Fall semesters had three sections each, with only one section over the Summer.  Students were 21.0\% female, 34.7\% non-white identifying, 56.3\% CS majors, and 31.0\% first-generation college students.  Final grades for the course over this time period were: 31.0\% As, 19.0\% Bs, 9.7\% Cs, 7.0\% Ds, 19.0\% Fs, and 14.3\% Withdrawn.

\subsection{Classroom Setting}
\label{subsection:classroom}
\subsubsection{Lecture}
All sections had the same instructor during lectures. Lecture sessions met twice per week for 1.5 hours (Spring, Fall) or 2 hours (Summer).  Each session consisted of a lecture, a lecture interspersed with in-class coding activities, a lecture with a quiz, or an exam, with the majority of classes containing in-class coding after the first week.  No changes were made to classwork, nor were any assignments added without credit for these experiments.

\subsubsection{In-Class Coding Activities}
In-class activities were recorded and monitored using GEM, a tool based on Code4Brownies \cite{Phan2018}.  GEM includes IDE plug-ins installed on the instructor's and students' computers.  It allows for exercises with starter code to be sent to students, student answers to be submitted to instructors and teaching assistants, and feedback/grades to be returned to students during class.  Submissions and grades are stored for later review.

In-class coding exercises were administered in the form of partially completed programs to help students focus on learning parts relevant to new topics. Students were given approximately 5 minutes to work on each simple problem and between 15 and 20 minutes for more complex problems. In-class activities were given one of three grades: 0 - for no submission or a lack of effort, 1 - for an incorrect answer, or 2 - for a correct answer.

\subsubsection{Lab}
Students attended a lab at the end of most weeks, where they were given a collection of programming exercises that were based on the current topic of that week's classes.
Topics include Java basics, variables, conditionals, loops, methods, arrays, and object-oriented programming (OOP).
Labs could be 0-2 days after the second lecture of the week for any given student, depending on which day of the week their lecture section met. Labs were taught over the course of 2 hours by 2 or 3 TAs and contained between 15 and 26 students from multiple class sections since students in all sections could freely choose which lab section to attend.
The last lab session consisted of the lab final, a comprehensive programming exam where students were not allowed to use any previous work, the internet, or notes of any kind.

\subsection{Data Processing and Collection}
As students solved the in-class coding assignments, GEM collected data on student performance and instructor feedback. Quiz, exam, and lab grades were manually collected by the instructor.  Since the point value of each lab assignment varied, we normalized them to a 0-100\% scale.

Anonymized data related to student performance was pulled from each section's in-class activities and divided into separate files.  Quiz, exam, and lab grades were then added for every student.

\subsection{Modeling Student Performance in CS1}
We experimented with popular machine learning methods to model ways of capturing students' eventual performance in CS1.
These methods entail training models to predict {\em target variables} based on {\em features}.
In the context of this work, we model students' eventual performance based on their performance on various assessment items during the course of a semester.

\subsubsection{Classification methods}
Machine learning methods we experimented with include support vector machines \cite{10.1007/978-3-642-03969-0_21}, logistic regression \cite{pregibon1981}, random forest \cite{Liaw2002}, K-nearest neighbors \cite{abu2019effects}, and Gaussian Naive Bayes \cite{zheng2005comparative}.
We used implementations of these methods in the widely-used scikit-learn library \cite{scikit-learn}.
In most cases, default parameters were used to avoid aggressive over-optimization.

\subsubsection{Features}
\label{sect:feature_selection}
Features are student performance during the course of a semester on in-class coding exercises, lab exercises, homework assignments, quizzes, and exams.
Features were categorized by type as either in-class coding or traditional work, as well as by due date.
Feature sets were divided by type as in-class only, traditional only, and hybrid.
They were further divided by time ranges into groupings of weeks, each from the beginning of the semester until a specific academic week's end.

\subsubsection{Target Variables}
Target variables, which we aim to predict, capture student's eventual performance, i.e. their performance at the end of the semester.
We experimented with different non-binary target variables, which capture different aspects of eventual performance, namely:
\begin{itemize}
    \item {\bf Final Exam Grade}, which are letter grades A, B, C, D, and F.  These letter grades were converted from scores that students got on the final exam.  This grade reflects a student's overall knowledge of CS1.  We used the standard formula to convert scores to letter grades: A's are $\geq$ 90\%, B's are ($\geq$ 80\% and <90\%), C's are ($\geq$ 70\% and <80\%), D's are ($\geq$ 60\% and <70\%), F's are <60\%. 

    \item {\bf Lab Final Grade}, which are letter grades A, B, C, D, and F.  These letter grades were converted from scores that students got on the final lab programming assessment. In our CS1 course, the lab final grade assesses student programming skills.  It uses the same standard formula to convert scores to letter grades, as was used for Final Exam Grades.
     
    \item {\bf Final Course Grade}, which are letter grades A+, A, A-, B+, B, B-, C+, C, C-, D+, D, and F.  These letter grades were calculated by the instructor based on a student's performance on all assignments given over the entire course of the semester.
    
    \item {\bf General Performance}, which are Failing, Struggling, Passing, and Succeeding.  This variable is calculated based on final course grades as follows: Failing (grades <50\%), Struggling ($\geq$ 50\% and <70\%), Passing ($\geq$ 70\% and <90\%), and Succeeding ($\geq$ 90\%).  
\end{itemize}

\subsection{Assessment Of Models}
Assessing a model's performance involves three main aspects: (1) choosing an appropriate metric to measure the model's predictive ability, (2) validating the model by properly training and testing the model, and (3) comparing it to at least one baseline.

First, we use {\em accuracy} to measure models' performance.  Accuracy is defined as the total number of correct predictions divided by the total number of predictions.
Combined metrics such as precision and recall can be used as metrics when target variables are binary and their distributions are greatly imbalanced.
In our study, the target variables are non-binary (multiple classes) and the data is relatively balanced, as such the use of accuracy to measure performance seems to be an appropriate choice.
This usage is similar to the study of Mao et. al \cite{mao19}, in which accuracy was the preferred metric for comparing different machine-learning methods.

Second, we utilize Stratified K-Fold cross-validation (with K=5) \cite{10.1007/978-3-642-29216-3_74}, which samples training data according to the distribution of target variables.
K-Fold is a popular method for validating models.  For instance, Wang et. al \cite{wang20} and \cite{mao19} also used K-Fold with K=5 to validate their models.
In our study, the target variables are non-binary with non-uniform distributions, thus it was appropriate to employ Stratified K-Fold to train models with training data that respects the target variable distributions.
To properly compare models, they were all trained and tested on the same partitions in each iteration of the Stratified K-Fold cross-validation.

Third, we contrast all model performance against simple baselines to give a more complete perspective of how well these models perform.
In many studies in CS education we have come across, researchers tend not to contrast their models against a baseline. We think that this omission makes it difficult to assess the performance of a model; an accuracy of 0.80 may have different implications depending on whether the problem is hard or easy.
When it is impractical to compare two different approaches to the same study, we think it is important to contrast a chosen modeling approach to a baseline.
For completeness, we contrast all models against three popular baseline predictive strategies: (1) {\bf Uniform Guessing}, which makes predictions by randomly sampling data according to an equal weighting of each label in the training data;
(2) {\bf Stratified Guessing}, which makes predictions by randomly sampling data according to the class distribution of the training data;
(3) {\bf Most Frequent Guessing}, which  always predicts the most frequent class in the training set.

%% file: 4_result.tex
\section{Results}
\label{section:results}
\begin{figure*}[t]
  \centering
  \includegraphics[width=1\linewidth]{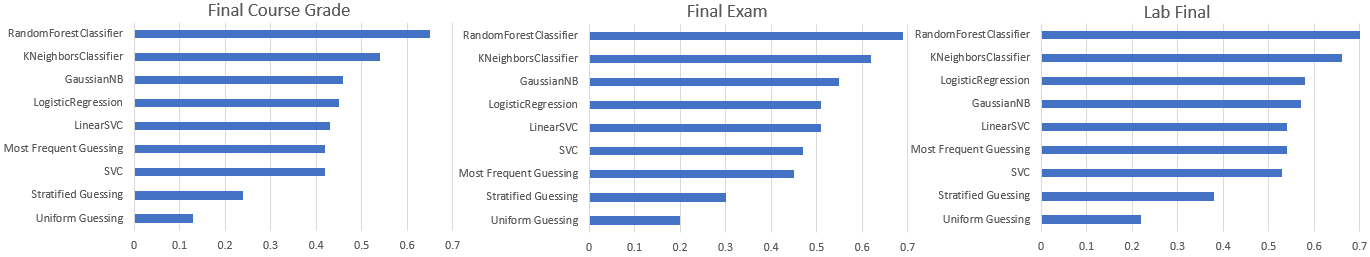}
  \caption{Accuracy in predicting Final Course Grade, Final Exam Grade, and Lab Final Grade for various models including baselines.}
  \label{fig:RQ0}
\end{figure*}

\subsection{Evaluating and Selecting Models}
\subsubsection{Establishing a Baseline Model}
To ensure that the predictive ability of the models is meaningful and nontrivial, we compare their performance to that of three baseline classifiers described in Section 3.5.
As seen in Figure \ref{fig:RQ0}, Most Frequent Guessing consistently outperformed the other two baselines by a wide margin and will thus be the baseline used for the rest of the paper.

\subsubsection{Selecting The Best Models}
We found that most classification methods had sufficient accuracy to be considered successful predictors compared to the best baseline model, see Figure \ref{fig:RQ0}.
The figure shows a comparison of accuracy between the models and all baselines, just before the final exams were given.
Final Course Grades are predicted with somewhat lower success than Final Exams and Lab Finals, the two most significant factors for determining Final Course Grades.

We found that Random Forest was the best-performing classifier across all targets under all conditions.
Then GridSearchCV was used to optimize Random Forest by exhaustively searching for 12 key parameters' (e.g. n\_estimators, max\_depth, and max\_features) optimal values across datasets from different semesters.
We will report results obtained from Random Forest with these optimal parameters.
When appropriate to contrast Random Forest's performance to a baseline model's, we will use results obtained from Most Frequent Guessing, which was the best baseline model observed.

\begin{figure*}[t]
  \centering
  \includegraphics[width=1\linewidth]{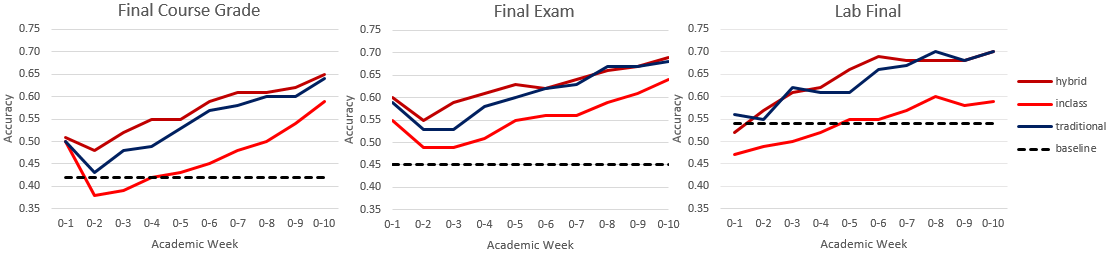}
  \caption{Accuracy of Random Forest Classification in predicting Final Course Grade, Final Exam Grade, and Lab Final Grade, from the beginning of the semester to the end of each academic week. In "Hybrid", both in-class exercises and out-of-class assignments and exams were used to build models.  In "Traditional", only out-of-class assignments and exams were used.}
  \label{fig:RQ1}
\end{figure*}

\subsection{RQ1: Can in-class exercises enhance the ability to predict students' eventual performance?}
\subsubsection{Setup}
For this experiment, we predicted eventual student performance for three slightly different aspects: final course grade, final exam grade, and lab final grade.
Lab final performance shows a student's cumulative ability to program and final exam performance reflects a student's eventual comprehension of the subject manner, whereas final course grade displays their overall performance throughout the course.
In terms of final course grades, at the end of a semester, the instructor calculates an overall score for each student.  These scores are converted to letter grades using the instructor's own scale (See Section 3.4.3).
In terms of final exam grades and lab final grades, since the instructor did not provide letter grades, we used the standard letter grade scale to convert final exam points to letter grades for model training (See Section 3.4.3).

In this experiment, we aim to compare prediction performance between three sets of features.
The first set of features (traditional) consists of student performance on homework assignments, lab assignments, quizzes, and exams.
These are the features most commonly used to predict students' grades in CS1.
The second set of features (inclass) is solely made of student performance on in-class coding assignments.
The third set of features (hybrid) contains all features from the traditional set as well as student performance on in-class coding assignments.

We hypothesized that the addition of in-class exercise features would improve our ability to predict final grades.
The results for final course grades are shown in Figure \ref{fig:RQ1}, which compares prediction performance over our three sets of features plotted against a baseline accuracy of 0.42, its highest baseline performance, as seen in Figure \ref{fig:RQ0}.
The final exam grade result comparison is plotted against a baseline accuracy of 0.45, and the lab final grade result comparison is plotted against a baseline accuracy of 0.54.
Prediction performance was measured at several academic week intervals during the course of a semester.
Each academic week interval is a period from the beginning of a semester to the end of an academic week.
Disregarding holidays, school closings, and study and exam days, there were a total of 10 academic weeks during a semester.

\subsubsection{Final Course Grade Findings}
As seen in Figure \ref{fig:RQ1}, we found that incorporating in-class exercises into traditional feature sets consistently performed better than only using traditional or inclass exercises as features.
This is further shown in the hybrid set's smaller fall in the 2nd academic week and faster recovery thereafter.
It can also be seen that inclass exercises alone are inferior prediction features as they are at most 10\% less accurate throughout the semester.

Despite all feature sets missing a step at the end of the 2nd academic week, they trend upward until the end of the semester, thereby recouping what they lost (See Section 5.2).

\subsubsection{Final Exam Findings}
As seen in Figure \ref{fig:RQ1}, we found that incorporating in-class exercises into traditional feature sets generally performed better than only using traditional or inclass exercises as features.
The traditional and hybrid feature sets converge at the end of academic week 6 just after the midterm exam, and remain close for the rest of the semester, seemingly indicating the importance of late-term assignments relative to students' final exam grades.
Once again, it can be seen that inclass exercises alone are inferior prediction features as they are at most 10\% less accurate throughout the semester.

It is also of note that all features performed better when predicting final exam grades than final course grades, as mirrored in Figure \ref{fig:RQ0}.

\subsubsection{Lab Final Findings}
As seen in Figure \ref{fig:RQ1}, we found that incorporating in-class exercises into traditional feature sets was only beneficial after the 4th academic week.
As with Final Exam prediction, Lab Final prediction for traditional and hybrid sets attempts to converge in the last half of the semester, which further strengthens the idea that late term assignments are important in determining students' lab final performance.
This time inclass exercises alone are shown to be inferior prediction features as they are at most 15\% less accurate throughout the semester

\subsection{Is it more useful to predict a student's general performance instead of their exact eventual performance?}
\begin{figure}[t]
  \centering
  \includegraphics[width=1\linewidth]{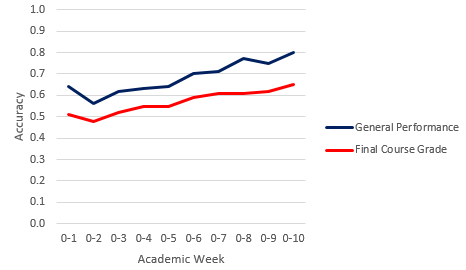}
  \caption{Accuracy of Random Forest in predicting an exact Final Course Grade versus a General Performance throughout the semester.}
  \label{fig:RQ2A}
\end{figure}

\subsubsection{Setup}
For this experiment, we made predictions on two slightly different aspects: eventual performance and general performance.
Eventual performance reflects our ability to predict a student's actual final course letter grade, whereas general performance reflects our ability to determine approximately how students are doing in a course.
Eventual performance is measured in relation to final course grades by using the instructor's own scale to calculate an overall score for each student at the end of a semester.
General performance is measured relative to a student's perceived standing in the course.
As predicting final course grades is more difficult than predicting either final exam or lab final grades, we will not be comparing them with general performance prediction.

In this experiment, we aim to compare prediction performance between two distinct target variables when predicting student performance on final course grades throughout the semester.
The first target variable FCG (Final Course Grade) consists of plus/minus letter grades as dictated by the course instructor (See Section 3.4.3).
The second target variable GP (General Performance) is composed of labels relating to how a student is handling the course (See Section 3.4.3).

We hypothesized that it would be easier to determine who to help by predicting students' general aptitudes than it is to predict student final grades.
The comparison of results can be seen in Figure \ref{fig:RQ2A}, which uses the same academic weeks mentioned at the end of Section 4.3.1.
For comparison's sake, Final Course Grade in Figure \ref{fig:RQ2A} uses the hybrid dataset as seen Figure \ref{fig:RQ1}.
No baseline is listed in Figure \ref{fig:RQ2A} to avoid confusion, as each target variable had a different best baseline score.

\subsubsection{Findings}
As shown in Figure \ref{fig:RQ2A}, it can be seen that predicting general performance is consistently more accurate than predicting a student's final letter grade throughout the course of the semester.
The lowest accuracy for GP (.56) is first beaten by FCG prediction halfway through the semester (.59, academic week 6).
Additionally, FCG prediction's highest accuracy (.65), which is achieved at the end of the semester, is only marginally better than GP prediction's initial accuracy at the start of the semester (.64).


As shown in Figure \ref{fig:RQ2A}, during academic weeks 3-5, accuracy remained relatively similar at about 0.62, immediately prior to the midterm exam.  This finding suggests that a general performance model can be useful in helping instructors identify which students need help, as early as the 3rd week.


\subsection{RQ2: To what extent does a curved final course grade impact prediction ability?}
\begin{figure}[t]
  \centering
  \includegraphics[width=1\linewidth]{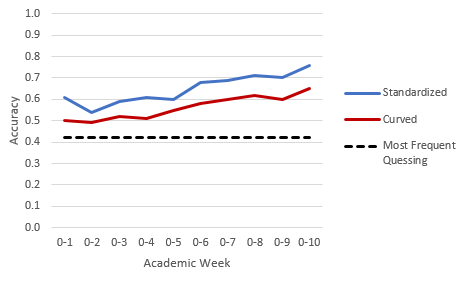}
  \caption{Standardized versus Curved Grading: Differences in Random Forest's prediction performance of final course grade using two different ways of assigning grades.}
  \label{fig:RQ3}
\end{figure}

 \subsubsection{Setup}
Although every course is different, exams and final grades are often curved \cite{kulick2008impact, dubey2010grading, tan2020students, czibor2014does, schinske2014teaching}.
These curves introduce biases that might make it difficult to predict student performance.
In this experiment, we sought to access the impact curving grades had on the prediction of eventual performance.

To contrast predictive results obtained from a specific instructor with their own subjective grading to those obtained from a standard grade scale, we compared the prediction performance of final course grades, which are determined in two different ways:

In the first way (Curved), final course grades were determined by a curve, specified by the instructor.  In this case, grades were A+, A, A-, B+, B, B-, C+, C, C-, D+, D, and F.
In the second way (Standardized), final course grades were determined using a standard scale of A, B, C, D, and F.

 \subsubsection{Findings}
As shown in Figure \ref{fig:RQ3}, we were able to predict final course grades more accurately when they were determined by the standardized scale than when they were determined by the instructor's curve.
This finding is understandable as the instructor's curve introduced biases to the grading in ways that the model was not able to accurately account for.
Nevertheless, predictive results had the same trajectories throughout the semester, with minor exceptions at the end of the 4th and 5th weeks.

%% file: 6_conclusion.tex
\section{Conclusion}
\label{section:conclusion}
Traditionally, in-class exercises give students opportunities to practice.
We demonstrated that they can also be helpful in improving the accuracy of predicting student performance.
Training models that incorporate student performance data from both in-class exercises and  traditional assignments yielded higher accuracy than training models that rely solely on traditional assignments.
We showed that academic weeks 3 to 5 were a good time to make predictions, which suggests that this approach can help formulate effective intervention schemes.